\documentclass[aps,preprint,showpacs,preprintnumbers,amsmath,amssymb]{revtex4}
\usepackage{amsmath,mathrsfs,amsbsy,color,graphicx,bm,amsthm,amsfonts}
\usepackage{units}
\usepackage{bbm}
\usepackage{times}
\usepackage{dcolumn}
\usepackage{mathrsfs}
\usepackage{amsmath,amssymb,epsfig}
\usepackage{amsmath}
\newcommand{\udots}{\mathinner{\mskip1mu\raise1pt\vbox{\kern7pt\hbox{.}}
\mskip2mu\raise4pt\hbox{.}\mskip2mu\raise7pt\hbox{.}\mskip1mu}}
\begin{document}
\title{Genuine  N-partite entanglement in Schwarzschild-de Sitter black hole
spacetime }
\author{Shu-Min Wu\footnote{Email: smwu@lnnu.edu.cn}, Xiao-Wei Teng, Xiao-Li Huang\footnote{Email: huangxiaoli1982@foxmail.com}, Jianbo Lu\footnote{Email: lvjianbo819@163.com} }
\affiliation{ Department of Physics, Liaoning Normal University, Dalian 116029, China
}


\begin{abstract}
Complex quantum information tasks in a gravitational background require multipartite entanglement for effective processing. Therefore, it is necessary to investigate the properties of multipartite entanglement in a relativistic setting. In this paper, we study genuine N-partite entanglement of massless Dirac fields in the Schwarzschild-de Sitter (SdS) spacetime, characterized by the presence of a black hole event horizon (BEH) and a cosmological event horizon (CEH). We obtain the general analytical expression of genuine N-partite entanglement shared by $n$ observers near BEH and  $m$ ($n+m=N$) observers near CEH. It is shown that genuine  N-partite entanglement monotonically decreases with the decrease of the mass of the black hole, suggesting that the Hawking effect of the black hole destroys quantum entanglement. It is interesting to note that  genuine N-partite entanglement is a  non-monotonic function of the cosmological constant, meaning that the Hawking effect of  the expanding universe can enhance quantum entanglement. This result contrasts with multipartite entanglement in single-event horizon spacetime, offering a new perspective on the Hawking effect in multi-event horizon spacetime.
\end{abstract}

\vspace*{0.5cm}
 \pacs{04.70.Dy, 03.65.Ud,04.62.+v }
\maketitle
\section{Introduction}
Quantum entanglement plays an important role in numerous quantum information processing tasks, including quantum cryptography, quantum teleportation, and dense coding \cite{L1,L2,L3,L4}. Unlike the usual multipartite entangled state, a genuine multipartite entangled state cannot be separated into any bipartite partitions. Genuine multipartite entanglement offers advantages over usual entanglement in the key resources for  measurement-based quantum computing and high-precision metrology \cite{L5,L6}.
Understanding genuine multipartite entanglement in a relativistic framework is crucial, especially considering the inevitable influence of gravity on quantum entanglement in real-world environments.
It is important to note that investigations of genuine multipartite entanglement near the event horizon of the black hole have been mainly limited to asymptotically flat spacetime \cite{L8,L9,L10,L11,L12,L13,L14,L15,L16,L17,L18,L19,L20,L21}. To clarify, genuine multipartite entanglement has been mainly studied in the single-event horizon  spacetime. The research papers have illustrated that the relativistic effects in this spacetime lead to the degradation of genuine multipartite entanglement.

The de Sitter solution is widely recognized as the most straightforward solution derived from Einstein's field equations with a nonvanishing cosmological constant \cite{L22,L23,L24,L25,LL25}. The study of phenomena in asymptotically de Sitter spacetimes is imperative and holds significant interest, particularly in light of experimental evidence indicating the accelerating expansion of our universe \cite{L26,L27}.
In reality, the black hole is asymptotically de Sitter, rather than asymptotically flat.
A static, chargeless black hole is associated with the Schwarzschild-de Sitter (SdS) spacetime, characterized by the mass $M$ of the black hole and the cosmological constant $\Lambda$. Therefore, the SdS spacetime features both a black hole event horizon (BEH) and a cosmological event horizon (CEH), introducing two-temperature thermodynamics distinct from those observed in single-event horizon  spacetime \cite{L28,L29,L30,L31}.  In comparison to single-event horizon spacetime, studying quantum information in multi-event horizon spacetime is more realistic, especially the exploration of multipartite entanglement, which has been a gap in the research. In addition, as relativistic quantum information tasks grow in complexity, the utilization of multipartite entanglement becomes essential for their processing. Hence, studying the relativistic effects of the SdS spacetime on genuine N-partite entanglement is one of the motivations for our work. Another motivation for our work is better to understand the multi-event horizon SdS spacetime through genuine N-partite entanglement.

In this paper, we study the properties of genuine N-partite entanglement of Dirac fields  in SdS spacetime endowed with the BEH and the CEH. Our model comprises $N$ modes: (i) the $n$ ($n<N$) modes located at the BEH; (ii) the $m$ ($n+m=N$) modes situated at the CEH.
We will derive the analytical expression for genuine N-partite entanglement in multi-event horizon spacetime. We aim to investigate how the Hawking effect of the black hole and the Hawking effect of the expanding universe influence genuine N-partite entanglement. Additionally,
we will investigate how genuine N-partite entanglement depends on $n$ and $m$ in the context of the multi-event horizon spacetime. As we all know, the gravitational effect of the single-event horizon spacetime destroys genuine N-partite entanglement \cite{L9,L10,L11,L12,L13,L14,L15,L16,L17,L18,L19,L20,L21}. An intriguing question arises:  will the gravitational effect of the multi-event horizon spacetime increase genuine N-partite entanglement?

The structure of the paper is as follows. In Sec. II, we describe the quantization of Dirac field in SdS spacetime. In Sec. III,  we study  the influence of the Hawking effect on genuine N-partite entanglement in multi-event horizon spacetime. The last section is devoted to the summary.
\section{Quantization of Dirac field in SdS spacetime\label{GSCDGE}}
The SdS spacetime metric is the unique solution to Einstein's field equations, including a positive cosmological constant $\Lambda$ in (3+1)-spacetime dimensions \cite{L28}.
The metric of the SdS spacetime can be given as
\begin{equation}\label{w1}
\begin{aligned}
ds^{2}=-\bigg(1-\frac{2M}{r}-\frac{\Lambda r^{2}}{3}\bigg)dt^{2}+\bigg(1-\frac{2M}{r}-\frac{\Lambda r^{2}}{3}\bigg)^{-1}dr^{2}+r^{2}\big(d\theta^{2}+\sin^{2}\theta d\phi^{2}\big).
\end{aligned}
\end{equation}
We shall now introduce the horizon structure of the SdS spacetime, which is dependent on the cosmological constant $\Lambda$. For a critical value of $\Lambda_{\rm crit}=1/(9M^2)$,  the event horizon of the SdS spacetime does not exist and the corresponding solution is
represented as the naked singularity. In the range $0<\Lambda<\Lambda_{\rm crit}$,
the SdS spacetime has the black hole  event horizon (BEH), the cosmological event horizon (CEH), and the unphysical event horizon for $f(r)=1-\frac{2M}{r}-\frac{\Lambda r^{2}}{3}=0$ \cite{L32}.
In this paper, we only consider the scope $0<\Lambda<\Lambda_{\rm crit}$ along
with the lapse function $f(r)$ that takes the form as
\begin{eqnarray}\label{w2}
f(r)=\frac{\Lambda}{3r}(r_H-r)(r-r_C)(r+r_H+r_C).
\end{eqnarray}
Here, the expressions for the $r_H$ (BEH) and $r_C$ (CEH) in terms of the mass
of the black hole and the cosmological constant can be written as
\begin{equation}\label{w3}
\begin{aligned}
r_{H}=\frac{2}{\sqrt{\Lambda}}\cos\bigg[\frac{\pi+\arccos(3M\sqrt{\Lambda})}{3}\bigg], \quad
r_{C}=\frac{2}{\sqrt{\Lambda}}\cos\bigg[\frac{\arccos(3M\sqrt{\Lambda})-\pi}{3}\bigg].
\end{aligned}
\end{equation}
The surface gravities of the black hole and the expanding universe can both be denoted as
\begin{equation}\label{w4}
\begin{aligned}
\kappa_{H}=\frac{\Lambda (2r_{H}+r_{C})(r_{C}-r_{H})}{6r_{H}},\quad -\kappa_{C}=\frac{\Lambda (2r_{C}+r_{H})(r_{H}-r_{C})}{6r_{C}}.
\end{aligned}
\end{equation}
From the equation above,  it becomes apparent that the surface gravity of the expanding universe manifests as negative, attributed to the repulsive effects induced by $\Lambda>0$. Because of $r_{H}<r_{C}$, we obtain $\kappa_{H}>\kappa_{C}$, showing that the Hawking temperature of the expanding universe $T_C=\frac{\kappa_{C}}{2\pi}$ is smaller than
the Hawking temperature of the black hole $T_H=\frac{\kappa_{H}}{2\pi}$ \cite{L33,L34,L35}.

To obtain the metric in the Kruskal coordinates, we introduce the tortoise coordinate, undergoing  transforms as  $\mu=t-r_{\star}$ and $\nu=t+r_{\star}$,  wherein the tortoise coordinate is denoted by \begin{equation}\label{w5}
\begin{aligned}
r_{\star}=\frac{1}{2\kappa_{H}}\ln\big|\frac{r}{r_{H}}-1\big|-\frac{1}{2\kappa_{C}}\ln\big|1-\frac{r}{r_{C}}\big|
+\frac{1}{2\kappa_{U}}\ln\big|\frac{r}{r_{U}}-1\big|,
\end{aligned}
\end{equation}
with $r_{U}=-(r_{H}+r_{C})$ \cite{ZL35,ZZL35}. Here, $\kappa_{U}$ is the surface gravity of the unphysical horizon $r_{U}$.
We need two Kruskal coordinate patches to get the non-singular coordinate mapping  for the entire SdS spacetime manifold through analytical continuation. The Kruskal coordinates are found to be
\begin{equation}\label{w6}
\begin{aligned}
\bar{\mu}_{H}=-\frac{1}{\kappa_{H}}e^{-\kappa_{H}\mu}, \quad \bar{\nu}_{H}=\frac{1}{\kappa_{H}}e^{\kappa_{H}\nu}, \quad
\bar{\mu}_{C}=\frac{1}{\kappa_{C}}e^{\kappa_{C}\mu}, \quad
\bar{\nu}_{C}=-\frac{1}{\kappa_{C}}e^{-\kappa_{C}\nu}.
\end{aligned}
\end{equation}
Finally, the BEH and CEH description of the metric in terms of the Kruskal coordinate can be expressed as
\begin{equation}\label{w7}
\begin{aligned}
ds^{2}=-\frac{2M}{r}\big|1-\frac{r}{r_{C}}\big|^{1+\frac{\kappa_{H}}{\kappa_{C}}}\big(1+\frac{r}{r_{H}+r_{C}}\big)^{1-\frac{\kappa_{H}}{\kappa_{U}}}
d\bar{\mu}_{H}d\bar{\nu}_{H}+r^{2}\Omega_2^2,
\end{aligned}
\end{equation}
\begin{equation}\label{w8}
\begin{aligned}
ds^{2}=-\frac{2M}{r}\big|\frac{r}{r_{H}}-1\big|^{1+\frac{\kappa_{C}}{\kappa_{H}}}\big(1+\frac{r}{r_{H}+r_{C}}\big)^{1+\frac{\kappa_{C}}{\kappa_{U}}}
d\bar{\mu}_{C}d\bar{\nu}_{C}+r^{2}\Omega_2^2.
\end{aligned}
\end{equation}

\begin{figure}
\includegraphics[scale=0.9]{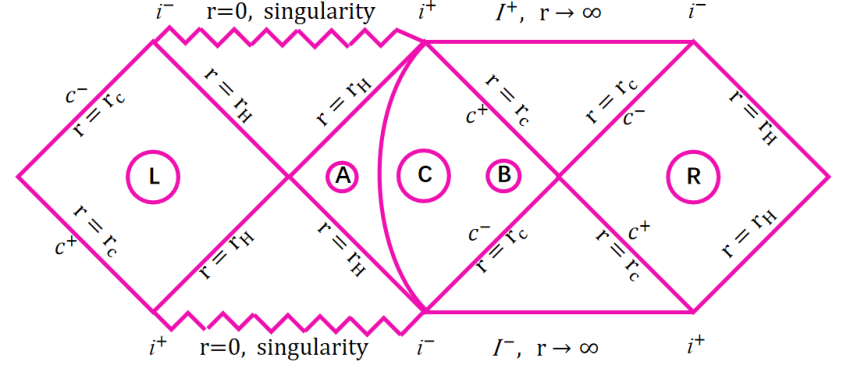}
\caption{ The SdS spacetime with thermal opaque membrane. }\label{Fig1}
\end{figure}

Upon review, it was discovered that the SdS spacetime has two physical event horizons associated with different local temperatures, meaning that they cannot reach thermal equilibrium.
In order to simplify the analysis, it is crucial to ensure that the system is in thermal equilibrium. The thermal opaque membrane serves this purpose. By employing it, it becomes possible to analyze one horizon with another as the boundary in the multi-event horizon SdS spacetime \cite{L36,L37}.
Therefore, the thermally opaque membrane divides region $C$ into two sub-regions, namely $A$ and $B$ ($C=A\cup B$) in Fig.\ref{Fig1}.  In our model, we consider that the $n$ observers located at the BEH can detect the Hawking radiation at temperature $T_H$, while the $m$ observers situated at the CEH can detect the Hawking radiation at temperature $T_C$.

The massless Dirac equation can be specifically expressed in the following form
\begin{eqnarray}\label{w9}
[\gamma^a e_a{}^\mu(\partial_\mu+\Gamma_\mu)]\Phi=0,
\end{eqnarray}
where $\gamma^a$ are the Dirac matrices and the four-vectors $e_a{}^\mu$
is the inverse of the tetrad $e^a{}_\mu$.
Let's first consider the sub-region $A$ and the causally disconnected region $L$, which
faces the BEH  in Fig.\ref{Fig1}.
In static de Sitter spacetime, normalizing field modes across the entire domain, from
$r=0$ to the CEH, presents challenges. Specifically, one needs to select mode functions that are analytic on the CEH, even though they may not be analytic at
$r=0$. This consideration becomes critical when discussing quantum entanglement \cite{GHJ1,GHJ2,L38,L39,L40,L41,L42}. Focusing on modes that are analytic at the CEH ensures that one can properly address the entanglement structure between different regions of spacetime, particularly across the CEH. While issues at $r=0$ could lead to potential singularities or ill-defined behavior, the primary interest in entanglement phenomena across the horizon justifies the focus on CEH-analytic modes. This approach allows for a consistent treatment of quantum correlations between spacetime regions separated by the horizon. The field quantization in the SdS spacetime can be performed in a similar way to the Unruh effect \cite{L38,L39,L40,L41,L42}.
Near the event horizon, when solving the Dirac equation, we obtain a set of positive-frequency (fermionic) outgoing solutions, distributed in regions both inside and outside the event horizon
\begin{eqnarray}\label{olmw1}
\Phi^+_{{\bold k},{\rm in}}\sim \phi(r) e^{i\omega u},
\end{eqnarray}
\begin{eqnarray}\label{olmw2}
\Phi^+_{{\bold k},{\rm out}}\sim \phi(r) e^{-i\omega u},
\end{eqnarray}
where $\phi(r)$ denotes four-component Dirac spinor and $u=t-r_{*}$. Here, $r_{*}\rightarrow \mp\infty$ as $r\rightarrow r_H$ and $r\rightarrow r_C$, respectively.
The Dirac field $\Phi$ can be expanded as
\begin{eqnarray}\label{olmw3}
\Phi&=&\int
d\bold k[\hat{a}^{\rm in}_{\bold k}\Phi^{+}_{{\bold k},\text{in}}
+\hat{b}^{\rm in\dag}_{-\bold k}
\Phi^{-}_{{\bold k},\text{in}}\nonumber\\ &+&\hat{a}^{\rm out}_{\bold k}\Phi^{+}_{{\bold k},\text{out}}
+\hat{b}^{\rm out\dag}_{-\bold k}\Phi^{-}_{{\bold k},\text{out}}],
\end{eqnarray}
where $\hat{a}^{\rm in}_{\bold k}$ and $\hat{b}^{\rm in\dag}_{-\bold k}$ are the annihilation and creation operators inside the event horizon, and $\hat{a}^{\rm out}_{\bold k}$ and $\hat{b}^{\rm out\dag}_{-\bold k}$ are the annihilation and creation operators outside the event horizon, respectively.
Unfortunately, for the SdS spacetime, there isn't a Kruskal coordinate that can simultaneously eliminate the coordinate singularities of both horizons. As a compromise, one needs to freeze the CEH to study genuine N-partite entanglement in the black hole region. Therefore, one can employ the Schwarzschild mode and the Kruskal mode for the quantization of the Dirac field, respectively, and then get the Bogoliubov transformations of the Schwarzschild and  Kruskal operators as \cite{L38,L42}
\begin{eqnarray}\label{olmw1}
\hat{c}^{\rm out}_{\bold k}&=&\cos r\hat{a}^{\rm out}_{\bold k}-\sin r\hat{b}^{\rm in\dag}_{-\bold k},\\
\hat{c}^{\rm out\dag}_{\bold k}&=&\cos r\hat{a}^{\rm out\dag}_{\bold k}-\sin r\hat{b}^{\rm in}_{-\bold k},
\end{eqnarray}
where $\cos r=\frac{1}{\sqrt{e^{-\frac{\omega}{T_H}}+1}}$, $\hat{c}^{\rm out}_{\bold k}$ and $\hat{c}^{\rm out\dag}_{\bold k}$ are the annihilation and creation operators acting on the Kruskal vacuum.
According to Bogoliubov transformations, the expressions for the Kruskal vacuum state and the excited state in the black hole spacetime are found to be
\begin{eqnarray}\label{w10}
\nonumber |0_{K_H}\rangle&=&\cos r |0_A,0_L\rangle+\sin r |1_A,1_L\rangle,\\
|1_{K_H}\rangle&=&|1_A,0_L\rangle,
\end{eqnarray}
with $\cos r=\frac{1}{\sqrt{e^{-\frac{\omega}{T_H}}+1}}$. Here, $|n_A\rangle$ and $|n_L\rangle$ denote the number states corresponding to the fermion outside the event horizon and the antifermion inside the event horizon of the black hole, respectively. Similarly, the expressions of the Kruskal vacuum state and the excited state in the expanding universe can be shown as
\begin{eqnarray}\label{w11}
\nonumber |0_{K_C}\rangle&=&\cos w |0_B,0_R\rangle+\sin w |1_B,1_R\rangle,\\
|1_{K_C}\rangle&=&|1_B,0_R\rangle,
\end{eqnarray}
 with $\cos w=\frac{1}{\sqrt{e^{-\frac{\omega}{T_C}}+1}}$.
Due to causal disconnection, the observer in the sub-region $A$ or $B$ cannot detect the modes of regions $L$ and $R$.
\section{Genuine N-partite entanglement in SdS spacetime\label{GSCDGE}}
If a state of the  N-partite system is not biseparable, it is named a genuinely N-partite entangled state. Here, we introduce the concurrence as a measure for genuine N-partite entanglement. The density matrix for the X-state of the N-partite system in the Hilbert-space orthonormal bases $\{|0,0,...,0\rangle,|0,0,...,1\rangle,...,|1,1,...,1\rangle\}$ can be expressed as
\begin{eqnarray}\label{w12}
 \rho_X= \left(\!\!\begin{array}{cccccccc}
\mathcal{M}_1 &  &  &  &  &  &  & \mathcal{C}_1\\
 & \mathcal{M}_2 &  &  &  &  & \mathcal{C}_2 & \\
 &  & \ddots &  &  &  \udots &  & \\
 &  &  & \mathcal{M}_y & \mathcal{C}_y &  &  & \\
 &  &  & \mathcal{C}_y^* &  \mathcal{N}_y & &  & \\
 &  & \udots &  &  & \ddots &  & \\
 & \mathcal{C}_2^* &  &  &  &  & \mathcal{N}_2 & \\
\mathcal{C}_1^* &  &  &  &  &  &  & \mathcal{N}_1
\end{array}\!\!\right),
\end{eqnarray}
with $y=2^{N-1}$. The conditions $\sum_i(\mathcal{M}_i+\mathcal{N}_i)=1$ and $|\mathcal{C}_i|\leq\sqrt{\mathcal{M}_i\mathcal{N}_i}$ show that the X-state $\rho_X$ is normalized and positive. Genuine N-partite concurrence can be denoted as
\begin{eqnarray}\label{w13}
C(\rho_X)=2\max \{0,|\mathcal{C}_i|-\mu_i \},   i=1,\ldots,y,
\end{eqnarray}
where $\mu_i=\sum_{j\neq i}^y\sqrt{\mathcal{M}_j\mathcal{N}_j}$ \cite{L43}.

In this paper, we initially consider an N-partite Greenberger-Horne-Zeilinger (GHZ) entangled state with $n$ Kruskal modes $\kappa_{H}$ and  $m$ Kruskal modes $\kappa_{C}$,
\begin{eqnarray}\label{w14}
|\psi\rangle_{1,\ldots,N}&=&\alpha|0_{\kappa_H}^1,0_{\kappa_H}^2,...,0_{\kappa_H}^n,0_{\kappa_C}^1,0_{\kappa_C}^2,...,0_{\kappa_C}^m
\rangle\\ \nonumber
&+&\sqrt{1-\alpha^2}|1_{\kappa_H}^1,1_{\kappa_H}^2,...,1_{\kappa_H}^n,1_{\kappa_C}^1,1_{\kappa_C}^2,...,1_{\kappa_C}^m
\rangle.
\end{eqnarray}
Here, $n$ ($0<n<N$) and $m$ ($0<m<N$) satisfy the relationship $n+m=N$.
To the Kruskal observers, the quantum system consists of $N$ modes.
Next, we let the $n$ modes be located at the BEH in the sub-region $A$ of $C$, and  the $m$ modes are located at the CEH in the sub-region $B$ of $C$.
Due to the Hawking effects of the SdS spacetime, the additional $n$ modes and $m$ modes appear in the region $L$ and the region $R$ in Fig.\ref{Fig1}, respectively. Using Eqs.(\ref{w10}) and (\ref{w11}), we can rewrite the initial state of Eq.(\ref{w14}) as
\begin{eqnarray}\label{w15}
|\psi\rangle_{1,\ldots,2N}&=&\alpha\bigg[ \bigotimes_{i=1}^{n}(\cos r |0^i_A,0^i_L\rangle+\sin r |1^i_A,1^i_L\rangle) \bigotimes_{j=1}^{m}(\cos w |0^j_B,0^j_R\rangle+\sin w |1^j_B,1^j_R\rangle) \bigg]\\ \nonumber
&+&\sqrt{1-\alpha^2}\bigg[ \bigotimes_{x=1}^{n}(|1^x_A,0^x_L\rangle) \bigotimes_{y=1}^{m}(|1^y_B,0^y_R\rangle)  \bigg].
\end{eqnarray}

Since the sub-regions $A$ and $B$ are causally disconnected from the sub-regions $L$ and $R$, we should take the trace over  physically inaccessible modes in the sub-regions L and R and then obtain the density operator $\rho_{N}$ as
\begin{eqnarray}\label{w16}
\rho_{N}=\rho_{\boldsymbol A}+\rho_{\boldsymbol X} +\rho^\dagger_{\boldsymbol X}+\rho_{\boldsymbol B},
\end{eqnarray}
with
\begin{eqnarray}
\nonumber\rho_{\boldsymbol A}&=&\alpha^2\bigg\{\bigg[\cos^{2n} r\bigotimes_{i=1}^{n}(|0\rangle^i_{A}\langle0|)\bigg]\bigg[\cos^{2m} w \bigotimes_{j=1}^{m}(|0\rangle^j_{B}\langle0|)\bigg]+\bigg[\cos^{2n} r\bigotimes_{i=1}^{n}(|0\rangle^i_{A}\langle0|)\bigg]\\\nonumber
&\times&\bigg[\cos^{2(m-1)} w \sin^{2} w \bigotimes_{j=1}^{m-1}(|0\rangle^j_{B}\langle0|)|1\rangle^m_{B}\langle1|\bigg]+...
\\\nonumber
&+&\bigg[\cos^2r\sin^{2(n-1)} r|0\rangle^1_{A}\langle0|\bigotimes_{i=2}^{n}(|1\rangle^i_{A}\langle1|)\bigg]\bigg[\cos^2 w \sin^{2(m-1)} w \bigotimes_{j=1}^{m-1}(|1\rangle^j_{B}\langle1|)|0\rangle^m_{B}\langle0|\bigg]\\\nonumber
&+&\bigg[\cos^2r\sin^{2(n-1)} r|0\rangle^1_{A}\langle0|\bigotimes_{i=2}^{n}(|1\rangle^i_{A}\langle1|)\bigg]\bigg[\sin^{2m} w \bigotimes_{j=1}^{m}(|1\rangle^j_{B}\langle1|)\bigg]\bigg]
\bigg\},\nonumber
\end{eqnarray}
\begin{eqnarray}\label{Q22}
\rho_{\boldsymbol X}=\alpha\sqrt{1-\alpha^2}\bigg[\cos^{n} r\bigotimes_{i=1}^{n}(|0\rangle^i_{A}\langle1|)\bigg]\bigg[\cos^{m} w \bigotimes_{j=1}^{m}(|0\rangle^j_{B}\langle1|)\bigg],\nonumber
\end{eqnarray}
and
\begin{eqnarray}
\nonumber\rho_{\boldsymbol B}&=&\alpha^2\bigg\{\bigg[\cos^{2(n-1)}r\sin^2r |1\rangle^1_{A}\langle1| \bigotimes_{i=2}^{n}(|0\rangle^i_{A}\langle0|)\bigg]\bigg[\cos^{2m} w \bigotimes_{j=1}^{m}(|0\rangle^j_{B}\langle0|)\bigg]\\\nonumber
&+&\bigg[\cos^{2(n-1)}r\sin^2r |1\rangle^1_{A}\langle1| \bigotimes_{i=2}^{n}(|0\rangle^i_{A}\langle0|)\bigg]\bigg[\cos^{2(m-1)} w \sin^{2} w \bigotimes_{j=1}^{m-1}(|0\rangle^j_{B}\langle0|)|1\rangle^m_{B}\langle1|\bigg]\\\nonumber
&+&...+\bigg[\sin^{2n} r\bigotimes_{i=1}^{n}(|1\rangle^i_{A}\langle1|)\bigg]\bigg[\cos^2 w \sin^{2(m-1)} w \bigotimes_{j=1}^{m-1}(|1\rangle^j_{B}\langle1|)|0\rangle^m_{B}\langle0|\bigg]\bigg\}\\\nonumber
&+&\bigg[\alpha^2\sin^{2n}r\sin^{2m}w+(1-\alpha^2)\bigg]
\bigg[\bigotimes_{i=1}^{n}(|1\rangle^i_{A}\langle1|)\bigg]\bigg[ \bigotimes_{j=1}^{m}(|1\rangle^j_{B}\langle1|)\bigg],\nonumber
\end{eqnarray}
which we write in matrix form as
\begin{eqnarray}\label{w17}
 \rho_{N}= \left(\!\!\begin{array}{cc}
 \mathcal{M}_{\boldsymbol A} & \mathcal{M}_{\boldsymbol X} \\
 \mathcal{M}_{\boldsymbol X}^T & \mathcal{M}_{\boldsymbol B} \\
 \end{array}\!\!\right),
\end{eqnarray}
in the $2^{N}$ basis
\begin{eqnarray}\label{Q22}
\nonumber&&\{|0^1_A,...,0^n_A,0^1_B,...0^m_B\rangle,|0^1_A,...,0^n_A,0^1_B,...0^{m-1}_B,1^m_B\rangle,...,
|1^1_A,...,1^n_A,1^1_B,...1^{m-1}_B,0^m_B\rangle,\\ \nonumber
&&|1^1_A,...,1^n_A,1^1_B,...1^m_B\rangle \}.\nonumber
\end{eqnarray}
The sub-matrixes $\mathcal{M}_{\boldsymbol A}$, $\mathcal{M}_{\boldsymbol X}$, and $\mathcal{M}_{\boldsymbol B}$ are detailed in Appendix A.

\begin{figure}
\begin{minipage}[t]{0.5\linewidth}
\centering
\includegraphics[width=3.0in,height=5.2cm]{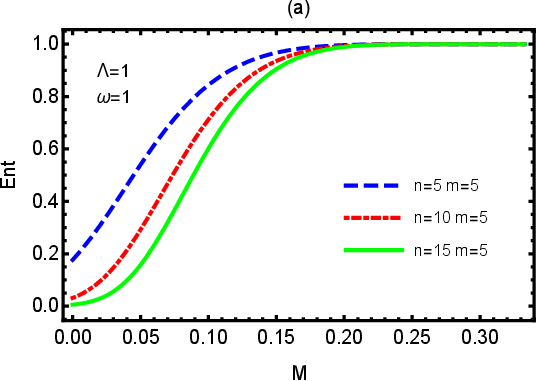}
\label{fig1a}
\end{minipage}%
\begin{minipage}[t]{0.5\linewidth}
\centering
\includegraphics[width=3.0in,height=5.2cm]{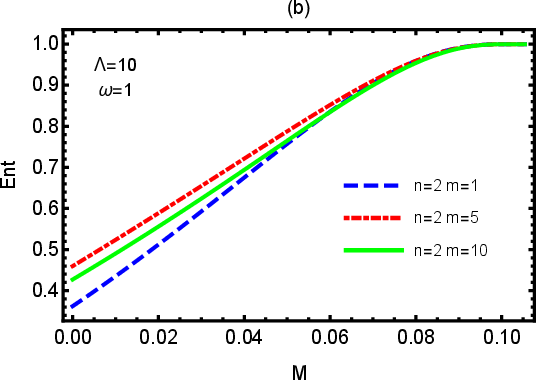}
\label{fig1c}
\end{minipage}%

\begin{minipage}[t]{0.5\linewidth}
\centering
\includegraphics[width=3.0in,height=5.2cm]{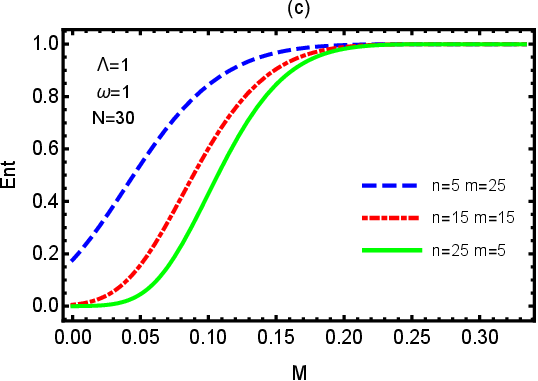}
\label{fig1a}
\end{minipage}%
\begin{minipage}[t]{0.5\linewidth}
\centering
\includegraphics[width=3.0in,height=5.2cm]{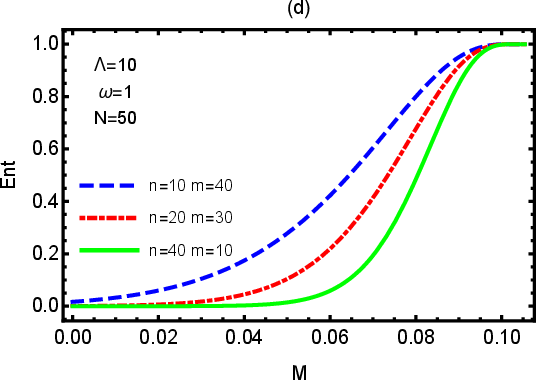}
\label{fig1c}
\end{minipage}%
\caption{Genuine N-partite entanglement $C(\rho_{N})$ as a function of the mass $M$ of the black hole for different $n$ and $m$, where  $\omega=\Lambda=1$.}
\label{Fig2}
\end{figure}

Employing Eqs.(\ref{w13}) and (\ref{w17}), we obtain the genuine N-partite entanglement measured by the concurrence as
\begin{eqnarray}\label{w18}
C(\rho_{N})=2\cos^nr\cos^mw\max \bigg\{0,\alpha\sqrt{1-\alpha^2}-(2^{N-1}-1)\alpha^2\sin^nr\sin^mw \bigg\}.
\end{eqnarray}
From Eq.(\ref{w18}), it is easy to see that genuine N-partite entanglement depends  not only on the initial parameters $\alpha$,  $n$, and $m$, but also on the mass $M$ of the black hole and the cosmological constant $\Lambda$.

In Fig.\ref{Fig2}, we plot genuine N-partite entanglement $C(\rho_{N})$ as a function of the mass $M$ of the black hole for different $n$ and $m$.
From Fig.\ref{Fig2}, we can observe that genuine N-partite entanglement decreases monotonically with the decrease of the mass $M$ of the black hole, meaning that the Hawking effect of the black hole degenerates quantum entanglement.
Note that genuine N-partite entanglement recovers to initial value ``$2\alpha\sqrt{1-\alpha^2}$" at the Nariai limit.  Note that
the degenerate solution where the black hole reaches its maximum size is called the Nariai limit. In this limit, the two horizons have the same size and temperature, so they are in thermal equilibrium. One can say that the energy the black hole loses due to evaporation is equal to the radiative energy it receives from the cosmological horizon. In recent papers \cite{L34,ZL35,L36,L37},  it is concluded that there is no gravitational influence on quantum entanglement in the considered limit case. Intuitively, any slight perturbation in the geometry could cause the black hole to become hotter than the background. Therefore, the thermal equilibrium of the Nariai limit is unstable. The black hole will be hotter than the cosmological horizon, and will therefore suffer a net loss of radiative energy. If neutral black holes spontaneously form in pairs within a de Sitter background, the initial conditions must satisfy the no-boundary condition, which imposes a specific linear combination of the two perturbation modes. Through the identification of suitable complex compact instanton solutions,  this condition results in black hole evaporation. Consequently, neutral primordial black holes are inherently unstable \cite{bn1}. This means that the Hawking effect of black holes has a greater impact on quantum entanglement than the Hawking effect of the expanding universe. We find that genuine N-partite entanglement monotonically decreases with increasing $n$, while  it exhibits non-monotonic changes  with increasing $m$.
Based on the Hawking temperature of the black hole $T_H$ being greater than the Hawking temperature of the expanding universe $T_C$, it can be concluded that genuine N-partite entanglement increases with the increase of $m$ for a fixed initial total number of particles $N$.  In other words,  genuine N-partite entanglement monotonically decreases as $n$ increases for a fixed $N$.
This conclusion is also supported by Fig.\ref{Fig2} (c) and (d).

\begin{figure}
\begin{minipage}[t]{0.5\linewidth}
\centering
\includegraphics[width=3.0in,height=5.2cm]{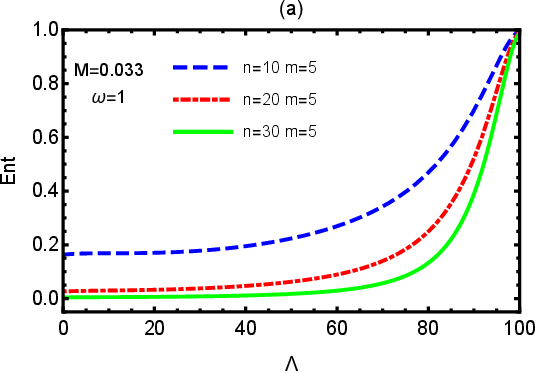}
\label{fig1a}
\end{minipage}%
\begin{minipage}[t]{0.5\linewidth}
\centering
\includegraphics[width=3.0in,height=5.2cm]{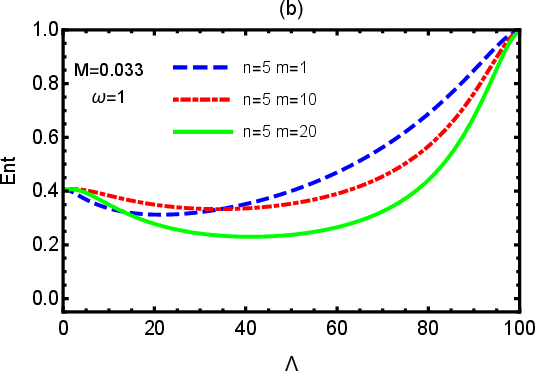}
\label{fig1c}
\end{minipage}%

\begin{minipage}[t]{0.5\linewidth}
\centering
\includegraphics[width=3.0in,height=5.2cm]{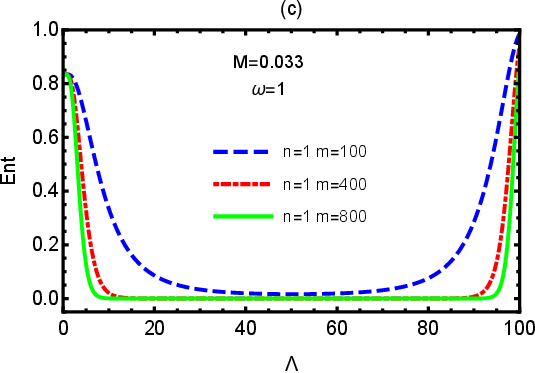}
\label{fig1a}
\end{minipage}%
\begin{minipage}[t]{0.5\linewidth}
\centering
\includegraphics[width=3.0in,height=5.2cm]{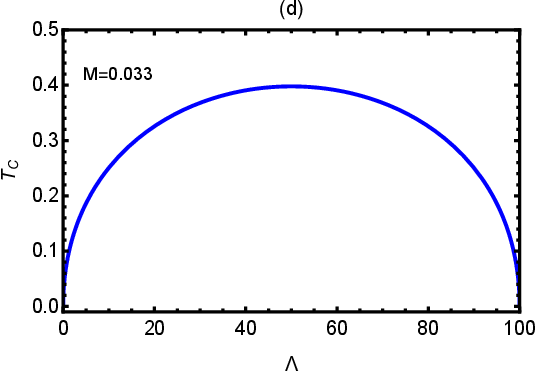}
\label{fig1c}
\end{minipage}%
\caption{Genuine N-partite entanglement $C(\rho_{N})$ and the Hawking temperature $T_C$ of CEH as functions of  the cosmological constant $\Lambda$ for different values of  $n$ and $m$, where $M=0.033$ and $\omega=1$.}
\label{Fig3}
\end{figure}

Fig.\ref{Fig3} (a)-(c) shows how the cosmological constant $\Lambda$  influences genuine N-partite entanglement $C(\rho_{N})$. From Fig.\ref{Fig3} (d), we see that
the Hawking temperature $T_C$ of CEH changes non-monotonically with $\Lambda$. We find that the Hawking effect of the expanding universe can enhance genuine N-partite entanglement, while the Hawking effect of the black hole only destroys genuine N-partite entanglement in the multi-event horizon spacetime. This results are different from the property of multipartite entanglement in the single-event horizon spacetime \cite{L9,L10,L11,L12,L13,L14,L15,L16,L17,L18,L19,L20,L21}. Fig.\ref{Fig3} (a)-(c) again shows that genuine N-partite entanglement is a decreasing function with $n$ and a non-monotonic function with $m$.

\section{ Conclutions  \label{GSCDGE}}
In this paper, we have studied the effect of the Hawking effect of the Schwarzschild-de Sitter (SdS) spacetime on genuine N-partite entanglement of massless Dirac fields shared by $n$ observers near BEH and $m$ ($n + m = N$) observers near CEH.
We obtain the general analytical expression of genuine N-partite entanglement for any $n$
and $m$ in the multi-event horizon spacetime. We find that the Hawking effect of the black hole can only degenerate genuine N-partite entanglement, while the Hawking effect of the expanding universe can enhance it. However, the Hawking effect of the single-event horizon spacetime destroys multipartite entanglement \cite{L9,L10,L11,L12,L13,L14,L15,L16,L17,L18,L19,L20,L21}.
This finding can contribute to a better understanding of the Hawking effect in multi-event horizon spacetime.
This is because the Hawking effect of the black holes and the Hawking effect of the expanding universe  have different influences on genuine N-partite entanglement.
Since the Hawking temperature of the black hole is bigger than
the Hawking temperature of the expanding universe, genuine N-partite entanglement increases with the increase of $m$  for a fixed initial parameter $N$. These conclusions demonstrate the observer-dependent nature of genuine N-partite entanglement in the multi-event horizon spacetime and guide multipartite entanglement to deal with relativistic quantum information tasks. On the other hand, we discuss how the effective temperature  affects genuine  N-partite entanglement in SdS spacetime (please see Appendix B for details.). In this context, particle creation at the effective temperature takes place within the region $C$, rather than in causally disconnected spacetime wedges. Our findings indicate that genuine N-partite entanglement  decreases steadily as the effective temperature rises. In contrast, genuine N-partite entanglement consistently increases with the Hawking temperature of the black hole. This is linked to the fact that the effective temperature rises with the black hole's mass. Thus, lowering the Hawking temperature actually leads to a higher effective temperature. Consequently, the distinct impacts of effective and Hawking temperatures on genuine N-partite entanglement arise from their fundamental differences.

\begin{acknowledgments}
This work is supported by the National Natural
Science Foundation of China (Grant Nos. 12205133, 12175095, and 12075050), the Special Fund for Basic Scientific Research of Provincial Universities in Liaoning under grant NO. LS2024Q002, LJKQZ20222315 and JYTMS20231051, and LiaoNing Revitalization Talents Program
(XLYC2007047).	
\end{acknowledgments}


\appendix
\onecolumngrid
\section{Sub-matrixes  $\mathcal{M}_{\boldsymbol A}$, $\mathcal{M}_{\boldsymbol X}$, and $\mathcal{M}_{\boldsymbol B}$ }
By analyzing  Eqs.(\ref{w16}) and (\ref{w17}), we find that the  matrix $\rho_N$ is $2^N\times2^N$ dimensions, and the sub-matrixes  $\mathcal{M}_{\boldsymbol A}$, $\mathcal{M}_{\boldsymbol X}$, and $\mathcal{M}_{\boldsymbol B}$ are $2^{N-1}\times2^{N-1}$ dimensions. First, the sub-density operator $\rho_{\boldsymbol A}$ corresponds to sub-matrix $\mathcal{M}_{\boldsymbol A}$ in the basis
\begin{eqnarray}\label{Q22}
\nonumber&&\{|0^1_A,...,0^n_A,0^1_B,...0^m_B\rangle,|0^1_A,...,0^n_A,0^1_B,...0^{m-1}_B,1^m_B\rangle,...,|0^1_A,1^2_A,...,1^n_A,1^1_B,...1^{m-1}_B,0^m_B\rangle,\\\nonumber
&&|0^1_A,1^2_A,...,1^n_A,1^1_B,...1^m_B\rangle \},
\end{eqnarray}
where the base corresponding to the element $\alpha^2\cos^{2(n-i)}r\sin^{2i}r\cos^{2(m-j)}w\sin^{2j}w$ include $i$ ``$1_A$" and $j$ ``$1_B$",
\begin{eqnarray}
\nonumber &&|0^1_A,...,0^n_A,0^1_B,...0^m_B\rangle\langle0^1_A,...,0^n_A,0^1_B,...0^m_B|: \alpha^2\cos^{2n}r\cos^{2m}w, \\\nonumber
&&|0^1_A,...,0^n_A,0^1_B,...0^{m-1}1^m_B\rangle\langle0^1_A,...,0^n_A,0^1_B,...0^{m-1}1^m_B|: \alpha^2\cos^{2n}r\cos^{2(m-1)}w\sin^2w,...,\\\nonumber
&&|0^1_A,1^2_A,...,1^n_A,1^1_B,...1^{m-1}_B,0^m_B\rangle\langle0^1_A,1^2_A,...,1^n_A,1^1_B,...1^{m-1}_B,0^m_B|:
\alpha^2\cos^2r\sin^{2(n-1)} r\cos^2w\times\\\nonumber
&& \sin^{2(m-1)}w, |0^1_A,1^2_A,...,1^n_A,1^1_B,...1^m_B\rangle\langle0^1_A,1^2_A,...,1^n_A,1^1_B,...1^m_B|:\alpha^2\cos^2r\sin^{2(n-1)} r\sin^{2m}w.
\end{eqnarray}
Therefore, the sub-matrix $\mathcal{M}_{\boldsymbol A}$ can be written as
\begin{eqnarray}\label{QQQQ24}
\mathcal{M}_{\boldsymbol A}=\alpha^2 \left(\!\!\begin{array}{cccc}
\cos^{2n}r\cos^{2m}w &  & &\\
  & \cos^{2n}r\cos^{2(m-1)}w\sin^2w& &\\
   &  & \ddots&\\
  &  & &\cos^2r\sin^{2(n-1)} r\sin^{2m}w \\
 \end{array}\!\!\right).
\end{eqnarray}
Second, the sub-density operator $\rho_{\boldsymbol X}$ corresponds to sub-matrix $\mathcal{M}_{\boldsymbol X}$ that can be easily written as
\begin{eqnarray}\label{q1}
\mathcal{M}_{\boldsymbol X}=\alpha\sqrt{1-\alpha^2} \left(\!\!\begin{array}{cccc}
 &  & &\cos^nr\cos^mw\\
  & & 0&\\
   & \udots & &\\
  0&  & &\\
 \end{array}\!\!\right).
\end{eqnarray}
Finally, the sub-density operator $\rho_{\boldsymbol B}$ can be demonstrated as sub-matrix $\mathcal{M}_{\boldsymbol B}$ that corresponds to elements
\begin{eqnarray}
\nonumber &&|1^1,0^2_A,...,0^n_A,0^1_B,...,0^m_B\rangle\langle1^1,0^2_A,...,0^n_A,0^1_B,...,0^m_B|:
\rho_{2^{N-1}+1}\\ \nonumber
&&|1^1,0^2_A,...,0^n_A,0^1_B,...,0^{m-1}_B,1^m_B\rangle\langle1^1,0^2_A,...,0^n_A,0^1_B,...,0^{m-1}_B,1^m_B|:\rho_{2^{N-1}+2},..., \\ \nonumber
&&|1^1,...,1^n_A,1^1_B,...,1^{m-1}_B,0^m_B\rangle\langle1^1,...,1^n_A,1^1_B,...,1^{m-1}_B,0^m_B|:
\rho_{2^{N}-1},\\ \nonumber
&&|1^1_A,...,1^n_A,1^1_B,...1^m_B\rangle\langle1^1_A,...,1^n_A,1^1_B,...1^m_B|:\rho_{2^{N}},
\end{eqnarray}
where
\begin{eqnarray}
\nonumber &&\rho_{2^{N-1}+1}=\alpha^2\cos^{2(n-1)}r\sin^2r\cos^{2m}w, \rho_{2^{N-1}+2}=\alpha^2\cos^{2(n-1)}r\sin^2r\cos^{2(m-1)}w\sin^2w, \\ \nonumber
&&\rho_{2^{N}-1}=\alpha^2\sin^{2n}r\cos^2w\sin^{2(m-1)}w, \rho_{2^{N}}=\alpha^2\sin^{2n}r\sin^{2m}w+1-\alpha^2.
\end{eqnarray}
Then, the sub-matrix  $\mathcal{M}_{\boldsymbol B}$ can be expressed as
\begin{eqnarray}\label{q7}
\mathcal{M}_{\boldsymbol B}= \left(\!\!\begin{array}{ccccc}
\rho_{2^{N-1}+1} &  & & &\\
  & \rho_{2^{N-1}+2}& & &\\
   &  & \ddots&\\
  &  & &\rho_{2^{N}-1} &\\
   &  & & &\rho_{2^{N}}\\
 \end{array}\!\!\right).
\end{eqnarray}

\section{Genuine  N-partite entanglement under the effective temperature }
Since the surface gravity of the black hole, denoted as $\kappa_H$, is greater than that of an expanding universe, $\kappa_C$, the flux of outgoing particles emitted from the BEH exceeds the flux of particles moving inward from the CEH at any point where $r_H<r<r_C$. This results in an effective outward flux, leading to the evaporation of the black hole. The effective temperature $T_{\rm{eff}}$ is closely linked to these particle fluxes. It is well established that an increase in the Hawking temperature of the black hole diminishes quantum correlations at a single horizon \cite{L8,L9,L10,L11,L12,L13,L14,L15,L16,L17,L18,L19,L20,L21}. However, the impact of effective temperature on genuine  N-partite entanglement remains ambiguous in the context of multi-event horizon spacetime. Thus, this section will investigate how the effective equilibrium temperature $T_{\rm{eff}}$, associated with the Hawking temperatures $\frac{\kappa_{H}}{2\pi}$ and $\frac{\kappa_{C}}{2\pi}$, influences genuine N-partite entanglement \cite{L34}.

The effective equilibrium temperature $T_{\rm{eff}}$ is linked to an emission probability that corresponds to the particle flux on the BEH created by the CEH. This indicates that particle creation occurs within a single region rather than across causally disconnected spacetime regions. In simple terms, the surface gravity $\kappa_U$ of the unphysical horizon can be expressed as
\begin{equation}\label{rw26}
\frac{1}{ \kappa_U}=\frac{1}{\kappa_C}-\frac{1}{\kappa_H}.
\end{equation}
It is important to note that the presence of $\kappa_U$ ensures the existence of the effective equilibrium temperature $T_{\rm{eff}}$. From Eqs.(\ref{w4}) and (\ref{rw26}), we observe that $T_{\rm{eff}}$ approaches $0$ as $\Lambda$ approaches $0$. We can utilize the coordinates ($\mu, \nu$) and  ($\bar\mu, \bar\nu$) with $\bar{u}=-\frac{1}{\kappa_{U}}e^{-\kappa_{U}\mu}$ and $\bar{\nu}=\frac{1}{\kappa_{U}}e^{\kappa_{U}\nu}$ to establish a field quantization framework. Following the standard quantization procedure, we derive the vacuum state $|\bar 0\rangle$ and  excited state $|\bar 1\rangle$ in the ($\bar\mu, \bar\nu$) mode, given by
\begin{eqnarray}\label{qww10}
\nonumber |\bar 0\rangle&=&\cos \gamma |0,0\rangle+\sin \gamma |1,1\rangle,\\
|\bar 1\rangle&=&|1,0\rangle,
\end{eqnarray}
with $\cos \gamma=\frac{1}{\sqrt{e^{-\frac{\omega}{T_U}}+1}}$. The effective temperature is defined as $T_{\rm{eff}}=\frac{\kappa_U}{2\pi}$. It is important to highlight that the creation of entangled pairs occurs solely in the region $A\cup B$ where $r_H<r<r_C$. As a result, $|\bar 0\rangle$ should not be regarded as a counterpart to the global or Minkowski vacuum. Additionally, we highlight that the presence of the surface gravity $\kappa_U$ (as opposed to $\kappa_H$ or  $\kappa_C$) is crucial for the emergence of the effective temperature $T_{\rm{eff}}$ \cite{L34}.

\begin{figure}
\begin{minipage}[t]{0.5\linewidth}
\centering
\includegraphics[width=3.0in,height=5.2cm]{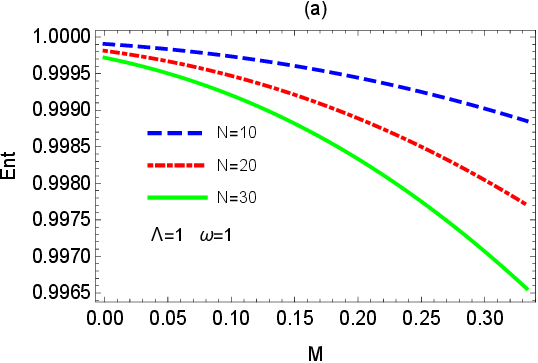}
\label{fig1a}
\end{minipage}%
\begin{minipage}[t]{0.5\linewidth}
\centering
\includegraphics[width=3.0in,height=5.2cm]{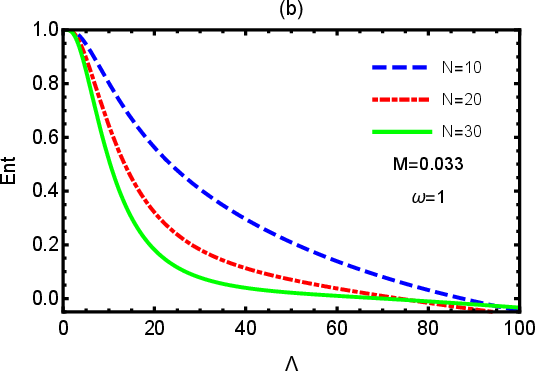}
\label{fig1c}
\end{minipage}%
\caption{Genuine N-partite entanglement $C(\bar\rho_{N})$ as a function of the mass $M$ of the black hole for fixed  $\Lambda=\omega=1$, and (b) as a function of the cosmological constant $\Lambda$ for fixed $M=0.033$ and $\omega=1$.}
\label{Fig4}
\end{figure}

We start by assuming that N-partite Greenberger-Horne-Zeilinger (GHZ) entangled state, as outlined in Eq.(\ref{w14}), resides in region $C$ and can be expanded using Eq.(\ref{qww10}). Through a series of detailed yet straightforward calculations, we derive the expression for genuine  N-partite entanglement as
\begin{eqnarray}\label{qww18}
C(\bar \rho_{N})=2\cos^N\gamma\max \bigg\{0,\alpha\sqrt{1-\alpha^2}-(2^{N-1}-1)\alpha^2\sin^N\gamma \bigg\}.
\end{eqnarray}
From Eq.(\ref{qww18}),  it is clear that the entanglement is influenced not only by the initial  parameters $\alpha$ and $N$, but also by the gravitational parameter $\gamma$.
Fig.\ref{Fig4} shows how the mass $M$ of the black hole and the cosmological constant $\Lambda$  influence genuine N-partite entanglement $C(\bar\rho_{N})$. It is shown that the entanglement $C(\bar\rho_{N})$ decreases monotonically with the increase of the $M$ and $\Lambda$. From Eq.(\ref{rw26}), we find that the effective equilibrium temperature $T_{\rm{eff}}$ rises consistently as both $M$ and $\Lambda$  increase. This indicates that, from the perspective of effective temperature, the entanglement $C(\bar\rho_{N})$ can be improved and destroyed  by the Hawking effect from the black hole and the Hawking effect from
the expanding spacetime, respectively.

\end{document}